\documentclass[12pt]{article}
\let\LARGE=\Large
\let\Large=\large
\let\large=\normalsize

\newcommand{\be}[3]{\begin{equation}  \label{#1#2#3}}     
\newcommand{\ee}{ \end{equation}}
\newcommand{\ba}{\begin{array}}
\newcommand{\ea}{\end{array}}
\newcommand{\bea}{\begin{eqnarray}}
\newcommand{\eea}{\end{eqnarray}}

\newcommand{\resetcounter}{\setcounter{equation}{0}}  

\newcommand{\ft}[2]{{\textstyle\frac{#1}{#2}}}

\def\beq{\begin{equation}}
\def\eeq{\end{equation}}
\def\beqa{\begin{eqnarray}}
\def\eeqa{\end{eqnarray}}

\usepackage{amssymb}

\begin{document}

\baselineskip=17pt
\parskip=8pt
\parindent=0pt


\thispagestyle{empty}

\begin{flushright}
\hfill{LMU-ASC 47/05}\\
\hfill{UUPHY/05-07}\\
\hfill{hep-th/0506251} \\
\hfill{\today}
\end{flushright}

\vspace{15pt}

\begin{center}{ \LARGE{\bf
Exploring  the relation between 4D and 5D BPS\\
\vskip 2mm
solutions
}}

\vspace{30pt}

{\bf Klaus Behrndt}$^a$, {\bf Gabriel Lopes Cardoso}$^a$
and {\bf Swapna Mahapatra}$^{b}$

\vspace{20pt}

$^a${\it Arnold-Sommerfeld-Center for Theoretical Physics \\
Department f\"ur Physik, Ludwig-Maximilians-Universit\"at
M\"unchen,\\
Theresienstra{\ss}e 37, 80333 M\"unchen, Germany
}\\[3mm]

{\tt behrndt@theorie.physik.uni-muenchen.de}\ ,
{\tt gcardoso@theorie.physik.uni-muenchen.de}

\vspace{10pt}

$^b${\it Physics Department, Utkal University,
Bhubaneswar 751 004, India} \\[1mm]
{\tt swapna@iopb.res.in}

\vspace{50pt}

{ABSTRACT}

\end{center}

\noindent
Based on recent proposals linking four and five-dimensional BPS
solutions, we discuss the explicit dictionary between general
stationary 4D and 5D supersymmetric solutions in $N=2$ supergravity
theories with cubic prepotentials.  All these solutions are completely
determined in terms of the same set of harmonic functions and the same
set of attractor equations. As an example, we discuss black holes and
black rings in G\"odel-Taub-NUT spacetime. Then we consider corrections
to the 4D solutions associated with more general prepotentials and
comment on analogous corrections on the 5D side.

\newpage


\section{Introduction and summary}

 
Supersymmetric solutions in five dimensions come in different
varieties.  Among the asymptotically flat solutions, there are
solutions describing black holes with or without rotation
 \cite{Tangherlini} -- \cite{9810204}
as well as black rings with rotation
\cite{0402144} -- \cite{0505167}. 
There is also a maximally supersymmetric solution describing a G\"odel
universe \cite{0209114}.  It is, moreover, known that a rotating black
hole or black ring in a G\"odel spacetime also yields a supersymmetric
solution \cite{0212002,0307194,0410252}.  In four dimensions, on the
other hand, there are supersymmetric multi-center solutions describing
a collection of extremal black holes
\cite{9705169} -- \cite{0304094}.

Recently \cite{0503217,0504126}, a very interesting relation has been
established between some of these five and four-dimensional
supersymmetric solutions.  Based on earlier remarkable work by
\cite{0209114}, it was shown in \cite{0503217,0504126} that
five-dimensional black holes and black rings, when embedded in a
Taub-NUT geometry, are connected to supersymmetric multi-center
solutions in four dimensions. This connection is implemented by using
the modulus of the Taub-NUT space to interpolate between the four and
the five-dimensional description.  In the vicinity of the NUT charge,
the spacetime looks five-dimensional, whereas far away from the NUT
the spacetime looks four-dimensional.  A black hole located at the NUT
will look like a five-dimensional black hole in the vicinity of the
NUT, whereas it will look like a four-dimensional black hole solution
far away from the NUT.  Similarly, a black ring sitting at some
distance from the NUT charge will, far away from the NUT, look like a
four-dimensional two-center solution.  One of these centers describes
the location of the NUT, whereas the other center describes the
position of the horizon of a four-dimensional supersymmetric black
hole.

The supersymmetric solutions of minimal supergravity in five
dimensions have been classified in the remarkable paper
\cite{0209114}.  In the case when the solution possesses a timelike
Killing vector, the solution is specified in terms of a hyper-K\"ahler
four-manifold describing the spatial base geometry orthogonal to the
orbits of the Killing vector field. If this base space admits a
tri-holomorphic Killing vector (i.e. a Killing vector which preserves
the hyper-K\"ahler structure), then the base space is a
Gibbons-Hawking space, and the full solution is determined in terms of
the Gibbons-Hawking metric and in terms of a set of harmonic
functions.  As described in \cite{0209114}, there is a dictionary
which relates a subset of these five-dimensional solutions to the
entire timelike class of supersymmetric solutions of four-dimensional
$N=2$ supergravity \cite{tod}.  In \cite{0504126}, this dictionary
between four and five-dimensional supersymmetric solutions has been
extended to general stationary solutions of $N=2$ supergravity
theories based on cubic prepotentials.

The relation established in \cite{0503217,0504126} implies that any
stationary five-dimensional solution of an $N=2$ supergravity theory,
when embedded in a Taub-NUT geometry, is connected to a
four-dimensional stationary solution of an associated four-dimensional
$N=2$ supergravity theory.  The four-dimensional solution is entirely
determined in terms of a set of harmonic functions and in terms of a
set of so-called stabilisation equations
\cite{9705169,9801081,0009234}. In the vicinity of a horizon these
equations are also known as attractor equations, and they were first
discussed in \cite{9508072,9602136,9603090}, and subsequently also in
\cite{9610105,9807056,9807087,9812082}.  It follows that also the
five-dimensional solution is entirely determined in terms of the same
set of harmonic functions and attractor equations.

In general, four-dimensional $N=2$ supergravity theories are not
simply based on cubic prepotentials, but on more general ones. This
implies that four-dimensional stationary solutions will be subject to
a variety of corrections, which can nevertheless be incorporated into
the solution in a systematic way thanks to the attractor mechanism
alluded to above.  If we now assume that the connection between
five-dimensional solutions in a Taub-NUT geometry and four-dimensional
solutions remains valid in the presence of corrections associated with
more general prepotentials, then we can determine the corrections to
five-dimensional quantities such as the five-dimensional entropy.
Evidence that this connection continues to hold in the presence of
$R^2$-corrections has recently been given in \cite{0505094}.

Let us explain the basic setup for connecting five to four-dimensional
solutions by considering a specific example, namely supersymmetric
Reissner-Nordstrom black holes in five and four dimensions. The
relation between four and five-dimensional solutions can be best
exhibited in a suitable coordinate system.

The line element of the five-dimensional supersymmetric
Reissner-Nordstrom black hole reads
\be230
ds^2 = -{1 \over H^2} dt^2 + H dx^m dx^m \quad , \qquad A = {dt \over H} \;,
\ee
where $H$ is a harmonic function given by
\be625
H = 1 + \Big({\rho_0 \over \rho} \Big)^2 \quad ,
\qquad dx^m dx^m = d\rho^2 + {\rho^2 \over 4} \Big[ \sigma_1^2 + \sigma_2^2
+\sigma_3^2 \Big] \;.
\ee
Here $\rho^2_0 = ({16 q^3 G_5^2 / \pi^2})^{1/3}$ \cite{9601029}, $q$
is the electric charge and $G_5$ denotes Newton's constant in five
dimensions.  The $\sigma_i, i=1,2,3,$ are the left-invariant
$SU(2)$-one forms given by
\bea
\sigma_1 &=& -\sin\psi d\theta + \cos\psi \sin\theta d\varphi \;, \nonumber\\
\sigma_2 &=& \cos\psi d\theta + \sin\psi \sin\theta d\varphi \;, \nonumber\\
\sigma_3 &=& d\psi + \cos\theta d\varphi \;,
\eea
where
$\theta \in [0, \pi]$, $\varphi \in [0,2\pi)$ and $\psi \in [0, 4\pi)$.

The entropy of this five-dimensional black hole
is
\be726
{\cal S}_5 = {A_5 \over 4 G_5} =  {2\pi^2 \rho^3_0 \over 4 G_5} =
{2 \pi} \sqrt{{q^3 }} \ .
\ee
Let us now relate this five-dimensional black hole solution to a
four-dimensional black hole solution.  To do so,
we perform the coordinate transformation
\be713
\rho^2 = 4 R r
\ee
and obtain for (\ref{625})
\be261
H = 1 +{\rho_0^2 \over 4 R r} \quad , \qquad
dx^m dx^m = {1 \over N} R^2 \sigma_3^2 + N \Big[dr^2 +
r^2 (\sigma_1^2 + \sigma_2^2)\Big]
\ee
with $N = {R /r}$. We note that since we have only done a coordinate
transformation, $H$ is still a localized harmonic
function. It is now straightforward to perform a reduction
over the compact coordinate $x^5 = R \psi$.  In doing so, one obtains
the line element for a four-dimensional supersymmetric black hole solution,
\be542
ds_4^2 = - {\rm e}^{2U} dt^2 + {\rm e}^{-2U} [ dr^2 + r^2 d\Omega_2]
\quad ,\qquad
{\rm e}^{-2U} = \sqrt{N H^3} \;,
\ee
which has the finite entropy
\be193
{\cal S}_4 = {A_4 \over 4 G_4} = {4 \pi \over 4 G_4}
\sqrt{ R \Big[{\rho_0^2 \over 4R}\Big]^3} =2 \pi \sqrt{q^3 } \;,
\ee
and we have used $G_5 = 4\pi R G_4$. Thus, we obtain an exact matching
of the entropies of the five and four-dimensional black hole, cf.\ eq.\
(\ref{726}).

Next, let us consider replacing the flat four-dimensional base space in
(\ref{261}) by a Taub-NUT base, namely
 \bea
  N = \frac{R}{r} \ \  \longrightarrow \ \ N = 1 + { p R \over r}
\; , \eea
where $p$ is the NUT charge.  The Taub-NUT space has a different
topology than four-dimensional flat space. It has a non-trivial $S_1$
associated with the angle $\psi$, whose asymptotic radius is finite and
given by $R$.  Near $r=0$, on the other hand, the space looks flat and
four-dimensional (for $p=1$).  We note that the pole of $H$ at $r=0$
is a NUT fixed point of the Killing vector $\partial_\psi$ and
therefore the black hole as given in (\ref{261}) is really localized
in all spatial directions and not smeared along $x^5$.

The entropy of the associated black hole, calculated
in either five or four dimensions, then becomes
\be621
{\cal S}_5 = {\cal S}_4 = 2 \pi \sqrt{p q^3 } \ .
\ee
{From} a four-dimensional point of view, this entropy exhibits the
expected quartic dependence on the charges.  The dependence on $p$
can be interpreted as a gravitational contribution to the entropy
\cite{9808085}.  More generally, one can also add angular momentum to
the five-dimensional black hole solution \cite{MyersPerry,9602065},
which will modify the result (\ref{621}).  Furthermore, one can also
take the four-dimensional base space in (\ref{261}) to be a
multi-center Gibbons-Hawking (GH) space \cite{GH}.

General $N=2$ supergravity theories are obtained by coupling an
arbitrary number of abelian vectormultiplets to $N=2$ supergravity.
Hypermultiplets can also be coupled to $N=2$ supergravity, but since
they play a spectator role in the context of stationary solutions,
they will not be considered in the following. The resulting
supersymmetric stationary five-dimensional solutions, when embedded in
a Taub-NUT geometry, are connected to supersymmetric stationary
four-dimensional solutions, as described above. The latter are
determined in terms of a set of harmonic functions and in terms of
so-called stabilisation equations.  Therefore also a five-dimensional
solution, which is connected to a four-dimensional solution, will be
determined in terms of the same set of harmonic functions and
stabilisation equations. The resulting dictionary will be discussed
in section 2, and can be summarized as follows.

The five-dimensional $N=2$ supergravity theory is based on the cubic
prepotential function $V (Y_{\rm 5d})= \frac16 C_{ABC} Y^A_{\rm 5d}
Y^B_{\rm 5d} Y^C_{\rm 5d}$, where the five-dimensional scalars
$Y^A_{\rm 5d}$ are real. The associated four-dimensional $N=2$
supergravity theory is based on the prepotential $F(Y) = - V(Y)/Y^0$,
where the four-dimensional scalars $Y^I = (Y^0, Y^A)$ are complex.  A
four-dimensional stationary solution is characterised by real harmonic
functions $ (H^I \ ; \ H_I) = (N, K^A \ ; \ M, L_A)$ associated to the
$Y^I$. The $Y^I$ are determined in terms of these harmonic functions
via the so-called stabilisation equations,
 \bea
Y^I - {\bar Y}^I = i H^I \;\;\;,\;\;\; F_I (Y) - {\bar F}_I (\bar Y)
= i H_I \;.
 \eea
Solving these equations one gets
\be141
Y^0 = {1 \over 2} (\phi^0 + i
\, N)\quad , \qquad Y^A = - {|Y^0| \over \sqrt{N}} \; x^A + {Y^0 \over
N} \; K^A  \;,
\ee
where
\be151
\frac12 C_{ABC} x^B x^C = L_A + \frac12\, {C_{ABC} K^B K^C
\over N} \, \equiv\, \Delta_A
\ee
and
 \bea
 \phi^0 = {\rm e}^{2 U} \left( N^2 M + N L_A K^A  + \frac13 C_{ABC} K^A
K^B K^C \right)\;.
 \eea
The four-dimensional line element reads $ds^2 = - {\rm e}^{2U} (dt +
{\vec{\omega}} \,d \vec{x})^2 + {\rm e}^{-2U} d \vec{x}^2$, with
 \bea
{\rm e}^{-4U} &=& {4 \over 9} N\, (x^A \Delta_A)^2 - N^{-2} \Big( N^2 M +
N {L_A K^A} + \frac13\, {C_{ABC} K^A K^B K^C } \Big)^2\ \;,
\nonumber\\
\nabla \times \vec{ \omega} &=&   N \nabla M -
M \nabla N + K^A \nabla L_A
- L_A \nabla K^A \;.
 \eea
The five-dimensional solution is entirely specified in terms of these
four-dimensional quantities.  The five-dimensional line element reads
$ds^2_5 = - f^2 (dt + \omega)^2 + f^{-1} ds^2_{GH}$, where $ds^2_{GH}
= N d\vec{x}^2 + R^2 N^{-1} (d \psi + {\vec A} d \vec{x} )^2 \;, \;
\nabla \times \vec A = R^{-1} \nabla N $ and $\omega = \omega_5 \, (d
\psi + {\vec A} d \vec{x} ) + \vec{\omega} d\vec{x}$.  We have the
following relations between the four and five-dimensional quantities,
 \bea
Y^A_{ \rm 5d} =  2^{1/3} x^A \;\;\;,\;\;\; f^{-3/2} = \frac23 \Delta_A x^A
\;\;\; , \;\;\; \omega_5 = R \, \frac{ {\rm e}^{- 2 U} \phi^0}{N^2}
\;. \label{5dstab}
\eea

This paper is organized as follows.  In section 2 we derive the
relation (\ref{5dstab}) which determines the five-dimensional
stationary solution in terms of the four-dimensional stabilisation
equations.  In section 3 we discuss specific examples, e.g.\ black
holes/black rings in G\"odel-Taub-NUT spacetime.  In section 4 we
focus on $R^2$-corrections to four-dimensional solutions and argue
that they will affect the five-dimensional solutions via the
four-dimensional stabilisation equations.  We consider, in particular,
the cloaking of three-charge solutions due to $R^2$-interactions in
four dimensions and we argue that a similar cloaking should occur for
two-charge solutions in a Taub-NUT geometry in five dimensions.  We
also comment on the recent discussion of higher derivative corrections
in four and five dimensions \cite{0505188,0505122}.

While this work was being finalized, two papers appeared on the
archive which have some overlap with ours. The paper \cite{0506222}
constructs supersymmetric black ring solutions in G\"odel spacetime
with the scalar fields valued in a symmetric space. This has some
overlap with our subsection 3.2 (however, we do not restrict the
scalar manifold to be a symmetric space).  The paper \cite{0506215}
discusses the cloaking of black rings, which has some overlap with
\mbox{section 4}.


\section{Relating four and five-dimensional BPS solutions}

\resetcounter


\subsection{Four-dimensional stationary BPS solutions}


Four-dimensional $N=2$ supergravity theories with $n$ abelian
vectormultiplets are based on a holomorphic prepotential function
$F(Y)$ and an associated symplectic section $(Y^I, F_I(Y))$, where
$F_I = \partial F(Y)/\partial Y^I$ ($I = 0, 1, \dots,n$), for more
details see \cite{deWitvanProeyen,Ferrara}.  Stationary supersymmetric
solutions in these theories are determined in terms of a set of
harmonic functions $(H^I, H_I)$.  The dependence of the scalar fields
$Y^I$ on these harmonic functions is determined via the so-called
stabilisation equations \cite{9705169,0009234}.  These were first
encountered when studying the entropy of supersymmetric black hole
solutions \cite{9508072,9602136,9603090}.  In the vicinity of the
black hole horizon, these equations are also called attractor
equations.  In the following, we will briefly review the form of a
stationary supersymmetric solution of these theories.

Any stationary line element can be written as
\be293
ds^2_4 = - {\rm e}^{2U} (dt + \vec{ \omega} \, d \vec{x})^2
+ {\rm e}^{-2U} d\vec x^2 \ .
\ee
The equations of motion for the gauge fields and the Bianchi identities are
solved in terms of a symplectic vector of harmonic functions in the three
coordinates $\vec x$,
\be612
(H^I \ ; \ H_I) = (N, K^A \ ; \ M, L_A) \;\;\;,\;\;\; A = 1, \dots, n \;.
\ee
These harmonic functions are, in general, multi-center functions which
are subject to a certain integrability condition (c.f. eq.
(\ref{382})).  Additional constraints may result by demanding absence
of closed timelike curves (CTCs), c.f. eqs. (\ref{334}), (\ref{339}).
The harmonic functions can, however, also be taken to be linear, in
which case they are not related to sources.  We will discuss various
choices of harmonic functions in the next section.

For the above stationary solution, the symplectic section $(Y^I, F_I)$
is determined in terms of the harmonic functions via the so-called
stabilisation equations,
\bea
Y^0 - \bar Y^0 = i \, N  & , \quad &Y^A - \bar Y^A = i \, K^A \;,\label{130}\\
F_0 - \bar F_0 = i \, M  & , &  F_A - \bar F_A = i \, L_A \;.\label{131}
\eea
The line element is determined by
\bea
\label{1313}
{\rm e}^{-2U} &=& i \, (\bar Y^I F_I - \bar F_I Y^I) \  , \\
\nabla \times \vec{ \omega} &=&   N \nabla M -
M \nabla N + K^A \nabla L_A
- L_A \nabla K^A
\label{729} \ .
\eea
Sources for the harmonic functions yield an integrability constraint
for equation (\ref{729}).  By contracting it with another derivative
one finds \cite{0005049},
\be382
0 = N \Delta M - M \Delta N +  K^A  \Delta L_A
- L_A \Delta K^A \ ,
\ee
where $\Delta$ is the three-dimensional Laplacian.  The
gauge field one-forms are given by \cite{0009234}
\be823
A^I = {\rm e}^{2U} (Y^I + \bar Y^I) \, (dt + \vec{ \omega} \, d\vec{x}
) - \, \vec{\alpha}^I d \vec{x}\;\;\;,\;\;\; \nabla \times
\vec{\alpha}^I = \nabla H^I
 \;.
\ee

In the following, we will consider stationary solutions based on the
cubic prepotential
\bea
F(Y) = \frac{D_{ABC} Y^A Y^B Y^C}{Y^0}  \;.
\label{cubicprepo}
\eea
Inserting this into (\ref{1313}) yields
\bea
{\rm e}^{-2 U} = |Y^0|^2 \, {\cal V}^3 \;,
\label{calvu}
\eea
where
\bea
{\cal V}^3 = -i D_{ABC} (z - \bar z)^A (z - {\bar z})^B (z - {\bar z})^C
\label{calv}
\eea
and $z^A = Y^A/Y^0$.

Next, we solve the stabilisation equations (\ref{130}) and (\ref{131})
for the cubic prepotential (\ref{cubicprepo}).  In doing so we will
closely follow \cite{9612076}.  The equations (\ref{130}) are solved
by
\be140
Y^0 = {1 \over 2} (\phi^0 + i \, N)\quad , \qquad
Y^A = - {|Y^0| \over \sqrt{N}} \; x^A + {Y^0 \over N} \; K^A \ .
\ee
The real quantities $x^A$ are determined via the stabilisation
equations $F_A - \bar
F_A = iL_A$, which read
\be150
-3 D_{ABC} x^B x^C = L_A - 3 \, {D_{A} \over N} \, \equiv\,  \Delta_A \;.
\ee
Here we introduced
\be821
D\equiv D_{ABC} K^A K^B K^C \ , \quad
D_A \equiv D_{ABC}  K^B K^C \ , \quad
D_{AB}\equiv D_{ABC}  K^C \ .
\ee
It follows that
\bea
z^A - {\bar z}^A = i \, \frac{\sqrt{N}}{|Y^0|} \; x^A \;\;\;,\;\;\;
{\cal V}^3 = - \frac{N^{3/2}}{|Y^0|^3} \, D_{ABC} x^A x^B x^C \;,
\label{zhp}
\eea
with $x^A$ completely determined in terms of the harmonic functions
(\ref{612}).  The $x^A$ are taken to be positive so that $T^A + 
{\bar T}^A = \sqrt{N} |Y^0|^{-1} x^A > 0$, where $T^A = - i z^A$.

The remaining stabilisation equation $F_0 - \bar F_0 = iM$ can then
be written as
\be160
N M = - \sqrt{N} D_{ABC} x^A x^B x^C \, {\phi^0 \over |Y^0|}
+ 3 D_{AB} x^A x^B  -  {D \over N} \;,
\ee
which yields
\be170 |Y^0|^2 = {N^3  (x^A \Delta_A)^2 \over 4 N (x^A \Delta_A )^2
- 9 \, N^{-2} \, (N^2 M + N {L_A K^A}  - 2 \, {D })^2 } \ee
and
\be721
{\rm e}^{-2U} \, \phi^0 =  N^2 M + N H_A K^A  - 2 D \ .
\ee
For the metric function (\ref{calvu}) one therefore obtains
\be629
{\rm e}^{-4U} = {4 \over 9} N\, (x^A \Delta_A)^2 - N^{-2}
\Big( N^2 M + {N L_A K^A}  - 2 \, {D } \Big)^2\ .
\ee
This completely determines the metric function ${\rm e}^{-2U}$ and the
scalar fields $z^A$ and $Y^0$ in terms of the set of harmonic
functions (\ref{612}).  The metric function ${\rm e}^{-2U}$ and the
scalar fields appear to be ill-defined when $N \rightarrow 0$, but
this is only an artifact of the parameterization in terms of the
quantities $x^A$.  This can be verified by performing a power series
expansion in $N$. This can also be directly checked for specific
examples.  For instance, consider the simple example of the so-called
STU-model, which is determined by the prepotential $F(Y) = - {Y^1 Y^2
Y^3 / Y^0}$. For this model, the metric function is obtained as
\cite{9608059,9608099}
\be927
\ba{l}
{\rm e}^{-4U} =  4 N L_1 L_2 L_3 - 4 M K^1 K^2 K^3 - (M N + L_1 K^1
+ L_2 K^2 +  L_3 K^3
)^2 \\
\hfill + 4 (L_1 K^1 L_2 K^2 + L_1 K^1 L_3 K^3 + L_2 K^2 L_3 K^3) \ .
\ea
\ee
The case discussed in the introduction (c.f. eq. (\ref{542})) is
obtained by equalizing the harmonic functions $L_A$ and setting $K^A =
M = 0$.


\subsection{Five-dimensional stationary BPS solutions}


Five-dimensional $N=2$ supergravity theories with $n$ abelian
vectormultiplets are based on a real cubic prepotential function
$V(X_{\rm 5d})$, where the $X^A_{\rm 5d}$ denote real scalar fields
($A= 1, \dots, n$), for more details see \cite{townsend}. These scalar
fields are constrained via
\bea
V (X_{\rm 5d}) = - D_{ABC} X^A_{\rm 5d} X^B_{\rm 5d} X^C_{\rm 5d} =
1 \;\;\;,\;\;
D_{ABC} = - \frac16 C_{ABC} \;.
\label{v5d}
\eea
Stationary supersymmetric solutions in these theories have been
constructed in the nice paper \cite{0408122}.  The five-dimensional
line element is given by
\bea
ds^2_5 &=& - f^2 (dt + \omega)^2  +
f^{-1} ds^2_{HK} \;, \label{5line}
\eea
where $ds^2_{HK}$ is the line element for any hyper-K\"ahler space. In
order to be able to connect five-dimensional solutions to the
four-dimensional solutions discussed in the previous subsection, we
take the hyper-K\"ahler space to admit a tri-holomorphic isometry, in
which case its line element can be written in terms of a
Gibbons-Hawking metric \cite{GH},
\bea
 \label{ghm}
 ds^2_{HK} &=& N d\vec{x}^2 + R^2 N^{-1} (d \psi +
\vec{A} \, d\vec{x} )^2
\;, \\
\nabla \times \vec{A}&=& R^{-1} \, \nabla N \;,
\label{nablatn}
\eea
where the modulus $R$ determines the radius of the $S^1$ circle
parameterized by the coordinate $\psi$.  The function $N$ can be any
harmonic function of $\vec x$.  For the Taub-NUT space, it is given by
\be002
 N = 1 + \frac{p^0 R}{r} \quad , \qquad
\psi \sim \psi + 4 \pi p^0 \;.
\ee
The periodicity in $\psi$ ensures the absence of conical singularities
(Dirac-Misner strings).

For a Gibbons-Hawking base space, the one-form $\omega$ in
(\ref{5line}) is given by \cite{0408122}
\be294
 \omega = \omega_5 \, (d \psi + \vec{A} d\vec{x} ) + \vec{
\omega} d \vec{x}\ ,
 \ee
 where
 \bea
 \label{omehat}
\nabla \times \vec{ \omega} &=&   N \nabla M - M \nabla N + K^A
\nabla L_A - L_A \nabla K^A \;,\\
\omega_5 &=& R \, N^{-2} \left( N^2 M  + N {L_A K^A} - 2 \, {D}
\right)\; . \label{omeg5}
 \eea
The metric function $f$, on the other hand, is determined as follows
\cite{9801161,0408122}. One introduces the rescaled variables
\bea
Y^A_{\rm 5d} = f^{-1/2} X^A_{\rm 5d} \;.
\eea
Then
 \bea
 V(Y_{\rm 5d}) = \frac16 C_{ABC} Y^A_{\rm 5d} Y^B_{\rm 5d}
Y^C_{\rm 5d} = f^{-3/2} \;. \label{fy}
 \eea
The $Y^A_{5d}$ are subject to the five-dimensional stabilisation
equations, as obtained in\footnote{The conventions used in
\cite{0408122} differ from ours in the following way: $L \rightarrow
{2^{2/3} \over 3} L \ , \ K \rightarrow - 2^{4/3} K$.} \cite{9801161,0408122}
\be299 \frac12
C_{ABC} Y^B_{5d} Y^C_{5d} = 2^{2/3} \left( L_A + \frac12\,
\frac{C_{ABC} K^B K^C}{N} \right) \;.
\ee
It follows that
 \bea
 f^{-3/2} = {2^{2/3} \over 3} \,\left( L_A + \frac12\,
\frac{C_{ABC} K^B K^C}{N} \right)
 Y^A_{\rm 5d} \ .
 \label{f32}
 \eea
Observe that, for the simple case given in (\ref{927}),
the function $f$ factorizes yielding
\be662
f^{-3} = 4 \, \Big( L_1 + \frac{K^2 K^3}{N}\Big)
\Big( L_2 + \frac{K^3 K^1}{N}\Big)
\Big( L_3 + \frac{K^1 K^2}{N}\Big) \;.
\ee
%


\subsection{Dictionary}


We are now in a position to discuss the dictionary between five and
four-dimensional stationary solutions. This dictionary has also been
given in \cite{0504126} using different conventions.

Comparing the five-dimensional stabilisation equations (\ref{299})
with the four-dimensional counterpart (\ref{150}), we see that they
are identical and therefore
\bea
\label{551}
Y^A_{\rm 5d} = 2^{1/3} x^A \;\;\;,\;\;\;
f^{-3/2} = {2 \over 3} \, \Delta_A
\, x^A \ .
\eea
 The equations for $\vec{\omega}$,  (\ref{omehat})
 and (\ref{729}), are also identical.  Comparing (\ref{omeg5})
 with (\ref{721}) yields the relation
\bea
 \omega_5 = R \, N^{-2} \,  {\rm e}^{- 2 U} \phi^0 \;.
\label{omeg5phi0}
 \eea
The $U(1)$ isometry of the Gibbons-Hawking metric (\ref{ghm}) is
generated by $\partial_\psi$ and hence we can perform the Kaluza-Klein
reduction over $\psi$ to four dimensions.  Following standard
formulae, we write the five-dimensional line element (\ref{5line}),
the five-dimensional gauge field one-forms and the five-dimensional
scalar fields as
\bea
\label{kkline5}
ds^2_5 &=& {\rm e}^{2 \phi} ds^2_4 + {\rm e}^{-4\phi}
(R \, d \psi - A^0_4)^2 \;,\\
A^A_{\rm 5d} &=& A^A_4 + {\rm Re} z^A \, (R \, d \psi - A^0_4)\;,\\
X^A_{\rm 5d} &=& - i \, {\rm e}^{2\phi} \, (z^A - {\bar z}^A) \;,
\label{sca}
\eea
where $ds_4^2$ denotes the four-dimensional line element given in
(\ref{293}), $(A^0_4, A^A_4)$ are the four-dimensional gauge field
one-forms and the $z^A$ denote the four-dimensional scalar
fields. Using (\ref{140}), the real part of $z^A$ is computed to give
\bea
z^A + {\bar z}^A &=&
i \, \frac{\phi^0}{N} \, (z^A - {\bar z}^A) + 2 \, \frac{K^A}{N}\ =\ -
{\phi^0 \over |Y^0|} \, { x^A \over \sqrt{N}} + 2 \, {K^A \over N} \;.
\label{z+z}
\eea
Comparing (\ref{kkline5}) with (\ref{5line})
we obtain
\bea
A^0_4 =  \frac{\omega_5}{ R } \, N^2 \, {\rm e}^{4U}\; (dt + \vec{\omega}\,
d \vec{x} ) -
R\,  \vec{A} d\vec{x}  \ .
\label{a0}
\eea
Using (\ref{omeg5phi0}) as well as (\ref{nablatn}) we see that (\ref{a0})
is in full agreement with (\ref{823}).  In addition, by
comparing (\ref{kkline5}) with (\ref{5line}) we also obtain
\bea
{\rm e}^{-4\phi}  = N^{-1} f^{-1} - \Big( \frac{f \, \omega_5}{R} \Big)^2
=
\frac{f^2}{N^2} \, {\rm e}^{- 4 U} \;.
\label{45cond}
\eea
Using (\ref{170}) and (\ref{551}) we find
\bea
|Y^0|^2 = {\rm e}^{4 U} \, N^3 \, f^{-3} \;,
\eea
and using (\ref{calvu}) we establish
\bea
{\rm e}^{-4\phi} =  {\cal V}^2 \;.
\label{vphi}
\eea
Using (\ref{551}), (\ref{omeg5phi0}) and (\ref{721}) it can be checked
that the expression for ${\rm e}^{-4U}$, as computed from (\ref{45cond}),
fully agrees with the expression (\ref{629}).  And finally, using
(\ref{170}), (\ref{629}), (\ref{calvu}) and (\ref{551}) we obtain
\bea
\frac{|Y^0|}{\sqrt{N}} = 2^{-1/3} f^{-1/2} {\cal V}^{-1} \;.
\eea
Together with (\ref{zhp}) and (\ref{551}) we establish
\bea
Y^A_{\rm 5d}&=  - i f^{-1/2} {\cal V}^{-1}\,  (z^A - {\bar z}^A) \;,
\eea
which is in precise agreement with (\ref{sca}).

Consistency of the solution requires the following positivity
constraints to be satisfied, namely
\bea
{\rm e}^{-4\phi} \ = \ {(Nf)^{-1} - \Big( {f \omega_5 \over R} \Big)^2}
\ = \ {f^2 \over N^2} \, {\rm e}^{-4U}  & > & 0 \ , \label{334} \\
\det ( -  {\rm e}^{2U} \omega_m  \omega_n + {\rm e}^{-2U} \delta_{mn})
\ = \ {\rm e}^{-2U}(-|\omega|^2 + {\rm e}^{-4U}) &>&0
\ , \label{339}
\eea
where $| \omega|^2 = \delta^{mn} {\omega}_m { \omega}_n$.  In
addition, for a given supergravity solution (for instance black
holes), one has to investigate whether Dirac-Misner strings are
present. Demanding their absence may enforce the additional constraint
$\vec{\omega} = 0$, at least near the centers of the solution and also
asymptotically.

To summarise, we see that the five-dimensional solution, which is
expressed in terms of $f, \omega_5, \vec{\omega}, Y^A_{\rm 5d}$ and
$A^A_{\rm 5d}$, is entirely expressed in terms of the harmonic
functions (\ref{612}) and in terms of the four-dimensional variables
$x^A$ and $Y^0$.  The latter are determined by solving the
four-dimensional stabilisation equations (\ref{130}) and (\ref{131}).
A related discussion on attractors and five-dimensional solutions has
appeared in \cite{0503219,0506177}.  Observe that, even though there
are $n$ five-dimensional real scalar fields $ Y^A_{\rm 5d}$, the
solution is expressed in terms of $2(n+1)$ harmonic functions
\cite{0408122}.

The four-dimensional stationary solutions are subject to a variety
of corrections associated to additional non-cubic terms in
the prepotential function (\ref{cubicprepo}). These corrections can be
computed in a systematic way thanks to the stabilisation equations,
which continue to hold \cite{0009234}. If, in the presence of these
corrections, the connection between four and five-dimensional solutions
in Taub-NUT geometries continues to hold, then the four-dimensional
stabilisation equations provide a powerful tool for computing corrections
to five-dimensional quantities, such as the entropy of a five-dimensional
black hole in a Taub-NUT geometry.


\section{Examples}

\resetcounter

Let us now discuss various examples in detail.
Each will correspond to a specific choice of the harmonic
functions introduced in (\ref{612}).


\subsection{Black holes and black rings and their entropy}


The simplest examples are provided by single-center black holes, which
are described by the harmonic functions
\bea
N = n + p^0 \, \frac{R}{ r} & , &
K^A = h^A + p^A\,  \frac{(R G_4)^{1/3}}{r} \;, \label{128} \\
M= m+ q_0 \, \frac{G_4}{R r} & , &
L_A = h_A+ q_A\,  \frac{(R G_4)^{2/3}}{R r} \ .
\label{129}
\eea
The integrability constraint (\ref{382}) becomes
(here we set $G_4 = R^2$)
\be722
m p^0  - nq_0  + h_A p^A  - h^A q_A
= 0\ .
\ee
The symplectic vector $(n, h^A ; m , h_A)$ comprising the constant
parameters and the symplectic charge vector $(p^0, p^A ; q_0 , q_A)$
are therefore mutually local.  In addition, one also has the
constraint ${\rm e}^{-2U} \rightarrow 1$ as $r \rightarrow \infty$.
Thus, there are two conditions on the constant parameters.  The number
of free parameters is therefore given by twice the number $n$ of
abelian vector multiplets.  In our conventions, the charges $(p^0, p^A
; q_0 , q_A)$ are integer valued and the dimensions are absorbed into
the factors of $G_4$ and $R$. That the charges $q_A$ and $p^0$ are
quantized in units of $(1/R)^{1/3}$ and $R$, respectively, is already
manifest in the example discussed in the introduction (c.f.
(\ref{193})). This fits with the general expectation that
electric/magnetic charges are associated with momentum/winding modes
along the circle in the $\psi$ direction.  The correct powers of $G_4$
and $R$ in $M$ and $K^A$ are then deduced from consistency.

{From} the four-dimensional point of view, all charges are on equal
footing and defined as asymptotic surface integrals, as usual. In five
dimensions, on the other hand, the $q_A$ are the usual electric
charges of the black hole, whereas the $p^A$ appear as dipole charges.
The charges $q_0$ and $p^0$ are on a different footing, namely $p^0$
is the NUT charge, whereas $q_0$ is related to the angular momentum of
the black hole.  The latter gets corrected by the other charges that
enter in $\omega_5$.

It is well known that there are CTCs hidden behind the
five-dimensional black hole horizon and that this solution becomes
pathological in the over-rotating case
 \cite{9810204,9906098,0003063,0211097,0211008,0302052,0308056}. 
In the
example given in (\ref{927}), the latter is manifest and occurs when
the function ${\rm e}^{-4U}$ becomes vanishing at the horizon.  This
happens when $q_0$ becomes large enough.  In four dimensions this
corresponds to a curvature singularity.  In five dimensions, on the
other hand, the function $f$ remains finite (since $x^A$ and
$\Delta_A$ are independent of $M$) (c.f.  (\ref{662})), but $\omega_5$
becomes large and renders the $\partial_\psi$- circle timelike
(c.f. (\ref{45cond}), (\ref{kkline5})).  At the point where the radius
of the circle vanishes, the four-dimensional solution is singular. A
consequence of the vanishing of ${\rm e}^{-2U}$ is that ${\cal V}$
also vanishes (c.f. \ref{vphi}) and hence, the scalar fields $z^A -
\bar z^A$ also go to zero.  This implies that the scalar fields are
deep in the interior of the K\"ahler cone.  In this regime instanton
corrections to the prepotential become relevant and they have the
property of regularising the solution and rendering the entropy
finite.  This has been discussed in \cite{9704095}.

For a generic choice of charges, the black hole has a regular horizon
and the entropy, calculated in the four-dimensional setting as well as
in the five-dimensional approach match exactly. If we denote the
two-dimensional horizon area in four dimensions by $A_4$, the entropy,
given by the Bekenstein-Hawking formula reads
\be622
{\cal S}_4  = {A_4 \over 4 G_4} = {4 \pi \over 4 G_4} \,
({\rm e}^{-2U} r^2)|_{r = 0} =  \pi \, {\rm e}^{-2U_0} \;,
\ee
where ${\rm e}^{-2U_0}$ is given by (\ref{629}), but with all
harmonic functions replaced by their quantized charges, i.e.\
$(N,K^A ; M,L_A) \rightarrow (p^0 , p^A ; q_0, q_A)$.  In five
dimensions the entropy is related to the three-dimensional area $A_5$ of the
horizon parameterized by the three coordinates $(\psi, \theta ,
\varphi)$.  Inspection of ({\ref{kkline5}),
(\ref{5line}) and (\ref{002}) shows that $A_5$
is given by
\bea
A_5 = 16 \pi^2 R p^0 {\rm e}^{- 2 \phi} f^{-1} N r^2|_{r =0}
= 16 \pi^2 R p^0 {\rm e}^{- 2 U_0} \;,
\eea
where we have used (\ref{45cond}).  The associated entropy is then given by
\be824
{\cal S}_5 = {A_5 \over 4 G_5} = \pi \, {\rm e}^{-2U_0} \ ,
\ee
where we used $G_5 = (4 \pi R p^0) \, G_4$. Hence \cite{0504126}
\bea
\label{442}
{\cal S}_5 = {\cal S}_4= 2 \pi \,
\sqrt{ p^0\,  \Big({\tilde x^A \tilde
\Delta_A \over 3 }\Big)^2 - (p^0)^2 \, J^2 }\ ,
\eea
where $\tilde x^A {\tilde \Delta}_A$ equals $x^A \Delta_A$ with the harmonic
functions replaced by the charges,
and
\bea
2 J = q_0 + \frac{p^A q_A}{p^0} - 2 \frac{D_{ABC} p^A p^B p^C}{(p^0)^2}
= \phi^0 {\rm e}^{-2U} N^{-2} R G_4^{-1} \, r|_{r =0} \;.
\eea
Observe that the pole in $p^0$
is only an artifact of the parameterization in terms of the $x^A$.

The single-center solution can be generalized to a multi-center one by
considering more general harmonic functions,
\bea
N = n + \sum_i  p^0_i  \, {R \over r_i} & , &
K^A = h^A + \sum_i \, p^A _i\,  \frac{(R G_4)^{1/3}}{r_i}  \;, \\
M= m + \sum_i \, q_0^i \, \frac{G_4}{R \, r_i}  & , &
L_A = h_A + \sum_i  \, q_A^i  \,  \frac{(R G_4)^{2/3}}{R \, r_i} \ ,
\label{122}
\eea
where $r_i = |\vec x - \vec x_i|$.  Inserting these functions into the
integrability constraint (\ref{382}) gives
\be320
\sum_i (N q_0^i - M p_i^0 + K^A q_A^i - L_A p^A_i )\,
\delta^{(3)}(\vec x - \vec x_i) = 0  \;,
\ee
here we have set $G_4 = R^2$ for simplicity.  By integrating these
equations without putting any constraints on the positions $\vec x_i$
of the centers, we obtain the following conditions
\bea
n q_0^j - m p^0_j + h^A q_A^j - h_A p^A_j  &= & 0 \label{412}
\qquad \forall \quad j \ , \\
 \, p^0_i q_0^j - q_0^i p^0_j + p^A_i q_A^j - q_A^i p^A_j \,
 &= & 0 \qquad \forall \quad i \neq j \label{443}  \ .
\eea
These conditions imply that the symplectic charge vectors $(p^0_i ,
p^A_i ; q_0^i , q_A^i)$ and the symplectic vector $(n , h^A ; m ,
h_A)$ are all mutually local. This severely constrains the parameters
and the charges. On the other hand, eq.\ (\ref{320}) can also be
seen as a constraint on the positions $\vec x_i$
\cite{0005049,0504126}. This gives the relation
\be822
N_i q_0^i - M^i p_i^0 + K^A_i q_A^i - L_A^i p^A_i  = 0 \;,
\ee
where $N_i \equiv N|_{{\vec x}=\vec{x}_i}$, $M^i \equiv M|_{\vec{x} =
\vec{x}_i}$, etc.  By
varying $\vec{x}_i$ one continuously changes the values of $N_i$, $M^i$,
etc, and hence these equations can always been solved.

For a generic choice of charges, each center describes a black hole,
from a four dimensional point of view.  {From} a five-dimensional
point of view, these centers may either correspond to black holes or
to black rings \cite{0504126}.

A particular four-dimensional two-center solution is connected to the
five-dimensional BPS black ring solution, which has attracted much
attention recently
\cite{0402144,0404073,0407065,0408010,0408120,0408122,0408186,0411187,
0504125,0504142,0505188,0506015}.  
This solution corresponds to the following choice of harmonic 
function\footnote{In order to simplify the notation we set $G_4 = R = 1$.}
\bea
K^A = \frac{p^A}{\Sigma}& , &  L_A = h_A + \frac{q_A} {\Sigma}\ ,
\nonumber\\
 M = - h_A p^A\, \Big(1- \frac{a}{\Sigma}\Big)
& , & N = n +   \frac{1}{r}\ ,  \nonumber\\
\Sigma =  |\vec x - \vec x_0|  &= &
\sqrt{r^2 +  a^2 + 2 r a \cos\theta} \;,
\label{399}
\eea
where $\vec x_0 = (0, 0, - a)$.
Therefore, the harmonic function $N$ is
sourced at the center $ r=0$, whereas the other harmonic functions
are sourced at the location of the black ring $\vec{x}_0$.

This choice of harmonic functions describes a black ring located at
$\vec{x}= \vec{x}_0$ with a horizon geometry $S^1 \times S^2$.  This
geometry is not (a deformed) $S^3$, because the Gibbons-Hawking fibre
is trivial at $\vec{x} = \vec{x}_0$. Hence one can always find a
coordinate system so that $d\psi + \vec{A} d\vec{x} = d\psi$ at the
position of the ring, which results in a factorized horizon geometry
(note that there is coordinate singularity at the horizon
\cite{0407065,0408122}).

In order to calculate $\vec{ \omega} \, d\vec{x}$ we can use the
expressions derived in \cite{0504142}, which in our notation become
\bea
\nabla \times \vec{\omega}^{(1)} &=& \nabla \, {1 \over r}
\quad \quad \quad \quad   {\rm with}  \quad
\vec{ \omega}^{(1)} \, d\vec{x}  = \cos\theta \, d\varphi \;, \label{220}  \\
\nabla \times \vec{ \omega}^{(2)} &=& \nabla \, {1 \over \Sigma}
\quad \quad \quad \quad \quad  {\rm with} \quad  
\vec{\omega}^{(2)} \, d \vec{x}
= {r \, \cos\theta + a \over \Sigma} \, d\varphi \;, \\
\nabla \times \vec{ \omega}^{(3)} &=& {1 \over \Sigma}\,  \nabla \, 
{1 \over r} - {1 \over r} \, \nabla \, {1 \over \Sigma}
\quad \quad   {\rm with} \quad
\vec{ \omega}^{(3)} \, d \vec{x}
= \Big( {r/a  +  \cos\theta \over  \Sigma} -{1 \over a} \Big) 
\, d\varphi \;, \label{221}
\eea
and hence, for the harmonic functions in (\ref{399}), $\vec{\omega} \,
d \vec{x}$ is given by
\be228
\vec{\omega} \, d\vec{x}
=   h_A p^A \, \Big( [\cos\theta + 1] \Big[ 1 - { a + r \over \Sigma}
\Big] \, + \, na \,  \Big[ { r\cos\theta + a \over \Sigma} -1 \Big] 
 \, \Big) \, d\varphi \;.
\ee
At $r = 0$, the quantities $M, \vec{\omega}$ and $\omega_5$ vanish.
The behavior at $r \rightarrow \infty$ depends crucially on the
Taub-NUT parameter $n$. If the constant part is not present, as for
the original black ring solution, $\vec \omega$ and $\omega_5$ vanish
asymptotically, but for $n \neq 0$, both quantities remain
finite. This raises the issue of the appearance of Dirac-Misner
strings, which can however be avoided if we choose the parameter $n$
in such a way that $\omega = \omega_5 (d\psi + \cos\theta d\varphi) +
\vec \omega d\vec x$ becomes trivial at infinity. On the other hand
the corresponding four-dimensional solution is still pathological
because $\vec \omega d\vec x \simeq \cos \theta d \varphi$ for $r
\rightarrow \infty$, and hence will have Dirac-Misner strings.  This
behavior may perhaps be avoided if one adds further appropriate
constant parts to the harmonic functions, for example to $M$.

The black ring solution corresponds to a two-center solution in four
dimensions, with the center at $\vec{x}_0$ describing a
four-dimensional black hole \cite{0504126}.  We can compute its
entropy by replacing the harmonic functions $(N, H^A; M, H_A)$ in
(\ref{629}) by the charges $(p^0 \, , p^A\, ,\, q_0\, ,\, q_A) = (1,
p^A, \, a h_A p^A \, , \, q_A)$.  For the example given in
(\ref{927}), we find
\bea
\label{733}
{\cal S}_{4} &=& \pi \left( {\rm e}^{- 2 U}
|\vec{x} -\vec{x}_0|^2 \right)|_{\vec{x} = \vec{x}_0}
\\
&=& 2 \pi \sqrt{
  (q_1 p^1 q_2 p^2 + q_1 p^1 q_3 p^3 + q_2 p^2 q_3 p^3)
- {(q_A p^A)^2 \over 4} - a \, (h_Ap^A) p^1 p^2 p^3
} \ , \nonumber
\eea
which is in agreement with the expression for the black ring
entropy given in \cite{0408122}.

The horizon of the black ring solution has geometry $S^1 \times S^2$
with an associated area of $2 \pi l$ and $ \pi \nu^2$, respectively
\cite{0408122}.  This horizon geometry is the same as the one of an
extremal BTZ black hole times a two-sphere \cite{0407065}.  The BTZ black hole
has entropy ${\cal S}_{3} =  \pi l/(2 G_3) $.  Using
 $G_3^{-1} = \pi \nu^2 G_5^{-1}$ gives ${\cal S}_3 = \pi^2
l \nu^2/(2 G_5)$, which is the entropy of the black ring \cite{0408122}.
On the other hand, the five-dimensional black ring in a
Taub-NUT geometry is connected to a four-dimensional black hole, as discussed
above.  We therefore have the equality
\bea
{\cal S}_{5} = {\cal S}_4 = {\cal S}_{3} \;
\eea
for the entropies.

One can, of course, also construct general multi-center solutions in
five dimensions \cite{0408122}.  A necessary condition for obtaining a
black ring instead of a black hole at a given center is the absence of
a source for $N$ at that point \cite{0504126}.  Upon reduction to four
dimensions all these solutions become multi-center black holes -- a
black ring can never become a single-center black hole.  {From} the
four-dimensional perspective, one can generate a five-dimensional
black ring by moving the entire NUT charge $p^0$ of a black hole to a
different position.  The original black hole is generically still
regular, but there is a naked singularity at the position of the NUT
charge.  In five dimensions this is a coordinate singularity, and this
process describes the topology change from $S^3 \rightarrow S^1\times
S^2$.


\subsection{Black holes and black rings in G\"odel-Taub-NUT spacetime}


A maximally supersymmetric G\"odel solution in five dimensions has
been obtained in \cite{0209114}.  Its metric function $f$ is constant
and the two-form $d\omega$ is anti-self-dual. This solution has CTCs
at every point in spacetime, similar to what happens for the
four-dimensional rotating G\"odel universe. CTCs at every point in
spacetime occur when a region of spacetime, where CTCs exist, is not
separated from the rest of spacetime by a black hole or a cosmological
horizon \cite{0401239}.  Various aspects of this supersymmetric
solution have been discussed in the literature
\cite{0212087,0301206,0309058,0404239,0405019}.

Supersymmetric solutions describing either a black hole or a black
ring in a G\"odel universe were constructed in
\cite{0212002,0307194,0410252} in minimal five-dimensional
supergravity.  Here we will construct black hole/black ring solutions
in G\"odel-Taub-NUT spacetime arising in five-dimensional supergravity
theories with abelian vectormultiplets.

The G\"odel solution of \cite{0209114} corresponds to the following
choice for the harmonic functions (\ref{612}),
 \be882
  M = {\cal G}
\, z = {\cal G} \, r \, \cos\theta \ , \qquad N = {R \over r} \ ,
\qquad L_A = h_A = {\rm const}\ ,  \qquad K^A = 0 \;,
 \ee
 with ${\cal G} = {\rm const}$.
 This ensures that the $Y^A_{\rm 5d}$ and $f$ are constant (c.f.
 (\ref{299}), (\ref{fy})).  The above choice of $N$ describes
 a flat four-dimensional base space.  The associated
 five-dimensional line element reads
\be727 ds_5^2 = - \Big( dt +  {\cal G} R\,r\,  [d\varphi +
\cos\theta \, d \psi]\Big)^2 + {R^2 \over N} (d\psi + \cos \theta d
\varphi)^2 + N \, ( dr^2 + r^2 d\Omega_2 ) \ . \ee
For large values of $r$ the timelike $U(1)$ fibration becomes
dominant, resulting in the appearance of CTCs, ie.\
$\partial_\varphi$ as well as $\partial_\psi$ are then inside the
future directed lightcone.

The set of harmonic functions in (\ref{882}) describing the G\"odel
deformation $\cal G$ can be superimposed with the set of harmonic
functions in (\ref{128}), (\ref{129}) and (\ref{399}).  The resulting solutions
then describe either a black hole or a black ring in a
G\"odel-Taub-NUT spacetime.

Let us first construct a black hole solution in a G\"odel-Taub-NUT
spacetime. Setting $G_4 = R^2$ for convenience, we consider the
following set of harmonic functions,
\bea
 N = {\cal G}_2\,  r \, \cos\theta + n + p^0 {R \over
r} & , &
K^A = h^A + p^A \frac{R}{r} \;, \nonumber\\
M= {\cal G}_1 \, r \, \cos\theta + m+ q_0 \frac{R}{r} & , & L_A =
h_A+ q_A \frac{R}{ r} \ ,
 \label{771}
  \eea
  where, for later convenience, we also allow for a G\"odel
  deformation of $N$ parameterized by ${\cal G}_2$. As in (\ref{220})
  -- (\ref{221}) we will first give the different contributions to
  $\vec\omega$ that involve a G\"odel deformation,
\bea
\nabla \times \vec{\omega}^{(4)} &=& \nabla \, (r \cos\theta)
\quad \qquad \qquad \qquad   {\rm with}  \quad
\vec{ \omega}^{(4)} \, d\vec{x}  = {1 \over 2} \, r^2 \sin^2\theta 
\, d\varphi \;,  \label{200}  \\
\nabla \times \vec{ \omega}^{(5)} &=& 
{1 \over r} \, \nabla \, (r \cos\theta) -
r \cos\theta \,  \nabla \,  {1 \over r} 
\quad  \quad {\rm with} \quad  
\vec{\omega}^{(5)} \, d \vec{x}
=  r \sin^2\theta \, d\varphi \;, \\
\nabla \times \vec{ \omega}^{(6)} &=& {1 \over \Sigma}\,  \nabla \, 
(r \cos\theta) - (a+{r \cos\theta}) \, \nabla \, {1 \over \Sigma}
\quad {\rm with} \quad
\vec{ \omega}^{(6)} \, d \vec{x}
= {r^2 \over \Sigma} \, \sin^2\theta 
\, d\varphi \;, \label{222}  
\eea
where $\Sigma = \sqrt{r^2 + a^2 +2 ra \cos\theta}$. 
With these expressions and the ones given in (\ref{220}) -- (\ref{221}) it is
straightforward to calculate $\vec\omega$ from (\ref{omehat}).  
This can actually be done for any two-center solution. If we
adjust the constants in the harmonic function in (\ref{771}) so that
$\vec\omega = 0$ for ${\cal G}_{1,2} = 0$, we obtain for a black hole
in a G\"odel-Taub-NUT spacetime
\bea
 {\vec \omega} \, d\vec{x}  \ = \ \Big[\,  {1 \over 2} (n {\cal G}_1 -
m {\cal G}_2 ) + (p^0 {\cal G}_1 
-q_0 {\cal G}_2 ) \, {R \over r}\,
\Big] \, r^2 \sin^2\theta  \, d\varphi \ .
 \label{881}
  \eea
  The remaining part of the solution, namely $\omega_5$ and $f$, is
  obtained by inserting the harmonic functions (\ref{771}) into
  (\ref{omeg5}) and (\ref{f32}).

Next, we construct a black ring solution in a
G\"odel-Taub-NUT spacetime. The black ring was described by the
harmonic functions in (\ref{399}). To the harmonic function $M$ we now
add the G\"odel deformation ${\cal G} r \cos \theta$, so that
\bea
K^A = \frac{p^A}{\Sigma}& , &  L_A = h_A + \frac{q_A} {\Sigma}\ ,
\nonumber\\
 M ={\cal G}  r \cos\theta - h_A p^A\, \Big(1- \frac{a}{\Sigma}\Big)
& , & N = n + \frac{1}{r}\ ,  \nonumber\\
 \Sigma = |\vec x - \vec x_0|  &= &
\sqrt{r^2 +  a^2 + 2 r a \cos\theta} \ .
\eea
In calculating  $\vec \omega$ we use the relations given 
(\ref{200}) --  (\ref{222}) as well as in (\ref{220}) -- (\ref{221}), 
and we obtain (here we set $R=1$)
\bea
\vec\omega d\vec x &= &\  {\cal G} \, \Big( \,  {n \over 2} \,  
 +  {p^0 \over r}\,
\Big) \, r^2 \sin^2\theta d\varphi
 \nonumber \\ &&
 +  h_A p^A \, \Big( [\cos\theta + 1] \Big[ 1 - { a + r \over \Sigma}
\Big] \, + \, na \,  \Big[ { r\cos\theta + a \over \Sigma} -1 \Big] 
 \, \Big) \, d\varphi \;.
\eea
Observe that in both cases the G\"odel deformation does not affect the
near horizon geometry, i.e.\ near the black hole at $r=0$ and near the
black ring at $r=a$, $\cos\theta = -1$ the G\"odel deformation either
vanishes or is constant. Therefore, the entropy of the black hole
remains unaffected. On the other hand, since $M$ grows linearly with
$r$, also $\omega_5$ grows with $r$ and we have to face the problem of
CTCs, as it happened in the overrotating case for black holes.  In
addition, also the four-dimensional solution can have CTCs, since the
condition (\ref{339}) will be violated with growing radial distance.
On the other hand, if one has two G\"odel deformations in $M$ and $N$
there is always a parameter choice so that the G\"odel deformations in
$\vec \omega$ cancel and CTC in four dimensions are avoided, which is
obvious in the expression (\ref{881}). More serious is the fact that
the four-dimensional solution exhibits a curvature singularity at some
finite radial distance which corresponds to the point where the circle
along the $\psi$ direction degenerates, i.e.\ where the condition
(\ref{334}) is violated and the solution becomes four-dimensional.  As
for the overrotating case, it would be interesting to discuss the
effect of instanton corrections or higher derivative corrections.
Observe that, with a growing harmonic function $M$, some of the scalar
fields become small and therefore, the simplest correction to the
prepotential in four dimensions which becomes important in this limit,
is the term $\sim i \chi \zeta(3) (Y^0)^2$, see \cite{9708065}, where
this term has been used as a regulator. Since $\chi$ is the Euler
number of the internal space, this term encodes some of the higher
derivative corrections in string theory. But before discussing effects
of higher derivative corrections in more detail, let us mention that
there is another (simple) possibility to avoid pathologies due to a
growing function $M$. Namely, replacing the G\"odel deformation in
(\ref{882}) with
\be660
   {\cal G} \, z \  \rightarrow \ {\cal G} \, ( 1 - |z -z_0|) \ ,
\ee
yields an upper bound when introducing a source at $z = z_0$, which
corresponds to a domain wall and is in the spirit of the discussion
in \cite{0404239}.  A generalization of this would be a periodic array
of sources yielding an upper and lower bound for the function $M$. In
doing so, one has however to keep in mind that these additional
sources also contribute to the integrability constraint (\ref{382}).


\section{Three-charge BPS solutions and $R^2$-corrections}

\resetcounter


Higher-order curvature corrections can convert an apparently
pathological solution of General Relativity into a regular solution
with an event horizon.  This so-called cloaking of a singularity has
recently been demonstrated to occur in string theory for certain
two-charge black hole solutions in four dimensions \cite{0409148,
0410076,0411255,0411272,0501014,0505122}.
One example of such a two-charge solution
is obtained in type IIA string theory on $K3 \times T_2$, by
wrapping $N_4$ D4-branes on $K3$ and adding a gas of $N_0$ D0-branes
to it. The resulting macroscopic entropy, which is entirely due to
higher-curvature terms in the effective action, is found to be given
by ${\cal S}_{\rm macro}=4 \pi \sqrt{N_0 \, N_4}$ in the limit of
large $N_0, N_4$. This is in agreement with a counting of the
microstates of the system \cite{0409148}.

The cloaking of singularities is not restricted to four dimensions.
As shown in \cite{0505122}, $R^2$-interactions in five (and higher)
dimensions can also cloak the singularity of two-charge solutions in
these dimensions.  In the following, we will use the recently
established connection between four and five-dimensional BPS
solutions \cite{0503217,0504126,0505094} to discuss the
cloaking of five-dimensional two-charge solutions in a Taub-NUT
geometry in terms of the cloaking of three-charge solutions in four
dimensions.  Here, the third charge is the Taub-NUT charge $p^0$,
which we take to be non-vanishing in order to be able to utilise the
connection between four and five-dimensional BPS solutions.

Analysing the cloaking of five-dimensional singularities in terms of
the four-dimensional solution has the advantage that in four
dimensions one can rely on a precise algorithm for constructing the
$R^2$-corrected BPS solution.  In five dimensions, on the other
hand, there is not yet a clear understanding of the nature of the
$R^2$-interactions and their impact on five-dimensional BPS
solutions.

$R^2$-interactions lead to a departure from the Bekenstein-Hawking
area law \cite{gr-qc/9307038} for the macroscopic entropy of a black
hole. In four dimensions, this departure is due to terms in the
effective Wilsonian action associated with the supersymmetrisation of
the square of the Weyl tensor \cite{9812082}.  On the other hand, the
departure from the area law in four and five dimensions has been
linked to a term in the effective action involving the Gauss-Bonnet
combination \cite{9711053,0505188,0505122}.  Thus, it would appear
that there are two combinations of $R^2$-terms giving rise to the same
leading correction to the entropy.  Here we will show that these two
combinations are actually equal to one another when evaluated on the
near-horizon solution. This may explain why the Gauss-Bonnet recipe
manages to reproduce some of the corrections to the macroscopic
entropy arising from a Wilsonian action with complicated
$R^2$-interactions.

Let us consider the near horizon geometry of a four-dimensional BPS
black hole solution.  This is a Bertotti-Robinson geometry, whose
static line element we write as $ds^2 = - {\rm e}^{2U} dt^2 + {\rm
e}^{-2U} d \vec{x}^2 $ with $U = \log r + {\rm const}$ and $r^2= x^m
x^m$.  This is a maximally supersymmetric solution of the equations of
motion of the Wilsonian $N=2$ Lagrangian with $R^2$-interactions.  Let
us evaluate the latter on this maximally supersymmetric solution. Most
of the terms in the Lagrangian vanish when evaluated on this maximally
supersymmetric background \cite{0009234}, and one is left with
\bea 8 \pi e^{-1} \, {\cal L}|_{\rm BR} = - \ft12 \, {\rm
e}^{- \cal K} \, R - \frac{i}{32} \left( F(X,{\hat A}) \, {\bar
{\hat A}}
 - {\rm h.c.} \right)
\;,
\label{lagn2sol}
\eea
where ${\hat A} = ( \varepsilon_{ij} T_{ab}^{ij} )^2$ and
$ {\rm e}^{\cal K}= G_{4}$ denotes Newton's constant in
four dimensions.

On
the solution, $ F(X,{\hat A}) \, {\bar {\hat A}} = {\rm e}^{4U} F(Y,
\Upsilon) \, {\bar \Upsilon}$, where $\Upsilon = {\bar \Upsilon} = -
64 \,U_m \,U_m$ and $U_m = \partial_m U$. 
Inserting this into (\ref{lagn2sol}) yields
\bea
8\pi e^{-1} \, {\cal L}|_{\rm BR} = - \ft12 \, {\rm e}^{- \cal K} \,
R - 4 \, {\rm Im} F(Y, \Upsilon) \, {\rm e}^{4U} \,U_m \, U_m \;.
\label{lagn2sol2}
\eea

The holomorphic function $F(Y, \Upsilon)$ has an expansion of the
form $F(Y, \Upsilon) = \sum_{g \geq 0} F^{(g)} (Y) {\Upsilon}^{g}$.
Here, $F^{(0)} (Y)$ denotes the prepotential function of subsection 2.1.
Let us now consider a particular function $F(Y,\Upsilon)$
of the form
\bea F(Y, \Upsilon) = F^{(0)} (Y) + F^{(1)}
(Y) \, \Upsilon \;,
\label{f1}
\eea
and let us rewrite the term proportional to $F^{(1)}$ in
(\ref{lagn2sol2}) in terms of the Gauss-Bonnet combination evaluated
on the solution. The Gauss-Bonnet combination $GB$ can be written as
$C^2 - 2 R_{\mu \nu}R^{\mu \nu} + \ft23 R^2$, where $C^2$ denotes the
square of the Weyl tensor.  The latter vanishes for conformally flat
solutions such as Bertotti-Robinson. Using $U_{mm} = U_m U_m = r^{-2}$
we note that $R = 2 (- U_{mm} + U_m U_m ) {\rm e}^{2U}$ also vanishes
(ignoring sources).  Using $R_{tt} = - U_{mm} {\rm e}^{4 U}$ and
$R_{mn} = - U_{pp} \, \delta_{mn} + 2 U_m U_n$, we obtain $R_{\mu
\nu}R^{\mu \nu} = 4 (U_m U_m)^2 {\rm e}^{4 U}$. Therefore, we find
that on the solution, (\ref{lagn2sol2}) can be written as\footnote{ In
heterotic string theory, $F^{(1)} = - i S /64 $ for large values of
the dilaton $S$ \cite{9906094}.  Inserting this into (\ref{lagn2sol3})
and using $G_{4}=2$ yields precise agreement with the heterotic
Lagrangian used in \cite{0505122} to compute the entropy of small
black holes.}
\bea
e^{-1} \, {\cal L}|_{\rm
BR} = - \frac{1}{16 \pi G_{\rm N}} \, R - \frac{1}{2 \pi}  \, {\rm
Im} F^{(0)}(Y) \, {\rm e}^{4U} \,U_m \, U_m - \frac{4}{\pi}  \, {\rm
Im} F^{(1)}(Y) \, GB \;.
\label{lagn2sol3}
\eea
Next we determine the correction to the Euclidean action due to the
term proportional to $F^{(1)}$ in (\ref{lagn2sol3}).  The Euclidean
solution is $H_2 \times S_2$ and has Euler character $\chi = (32
\pi^2)^{-1} \int GB = 1 \times 2= 2$. Using the fact that the scalar
fields $Y$ are constant in a Bertotti-Robinson spacetime, we find that
the $F^{(1)}$-term in (\ref{lagn2sol3}) contributes the following
amount to the Euclidean action,
\bea \Delta S_E = - 256 \pi {\rm Im} F^{(1)} (Y) \;.
\label{corraction}
 \eea
This we now compare
with the corrections to the macroscopic entropy formula due to
$R^2$-interactions.  The macroscopic entropy computed from the
effective Wilsonian Lagrangian is given by \cite{9812082} 
\bea {\cal
S}_{\rm macro} = \pi \left[ |Z|^2 - 256 {\rm Im} F_{\Upsilon} (Y,
\Upsilon)
 \right] \;,
\eea
where here $\Upsilon = -64$ and
$F_{\Upsilon} = \partial F/\partial \Upsilon$. For the function
(\ref{f1}) this gives
\bea
{\cal S}_{\rm macro} = \pi \left[ |Z|^2 - 256 {\rm Im}
F^{(1)} (Y)
 \right] \;.
\label{enrof1}
\eea
We therefore see that the correction to the Euclidean action
(\ref{corraction}) precisely equals the correction term proportional
to $F^{(1)}$ in the macroscopic entropy (\ref{enrof1}).  The latter is
the Wald term which measures the deviation from the area law of
Bekenstein and Hawking.

The above agreement suggests to view the Gauss-Bonnet recipe as an
effective recipe which manages to capture some of the corrections to
the entropy due to the complicated supersymmetrised $R^2$-terms.

Next, let us discuss the cloaking of three-charge solutions in four
dimensions. For convenience, we will consider solutions of heterotic
string theory on $K3 \times T_2$.  The four-dimensional
$R^2$-corrected effective Wilsonian action is known to contain a term
$(S + \bar S)^2 C_{\mu \nu \rho \sigma}^2$ at tree-level, where
$C_{\mu \nu \rho \sigma}$ denotes the Weyl tensor and $S$ the dilaton
field.  The tree-level holomorphic function $F(Y, \Upsilon)$
associated with a heterotic $N=2$ compactification on $K3 \times T_2$
is given by
\bea
 F(Y, \Upsilon) = - \frac{Y^1 Y^a \eta_{ab} Y^b}{Y^0} + c_1
\,\frac{Y^1}{Y^0}\, \Upsilon \;,
\label{prepo}
\eea
where we have suppressed instanton contributions.
Here 
\bea Y^a \eta_{ab} Y^b
= Y^2 Y^3 - \sum_{a = 4}^n (Y^a)^2 \;\;\;,\,\;\; a = 2, \dots, n \;,
\eea
with real constants $\eta_{ab} = \ft12 C_{ab}$ and $c_1 = -
\ft{1}{64}$.  The $C_{ab}$ denote the intersection numbers of $K3$.
The dilaton field is defined by $S = -i Y^1/Y^0$. The moduli $T^a$ are
given by $T^a = - i Y^a/Y^0$.

The Wilsonian $N=2$ Lagrangian based on the holomorphic function $F(Y,
\Upsilon)$ has supersymmetric charged multi-center solutions
\cite{0009234,0012232}.  The one-center solutions are static and
spherically symmetric.  The associated line element is given by $ds^2
= - {\rm e}^{2U} dt^2 + {\rm e}^{-2U} d \vec{x}^2$, where
\cite{0009234}
\bea
{\rm e}^{-2U} = i \left[ {\bar Y}^I \, F_I (Y, \Upsilon)
- {\bar F}_I ({\bar Y}, {\bar \Upsilon}) \, Y^I \right]
+ 128 i \, {\rm e}^U \,
\nabla^p \left[ {\rm e}^{-U} \, \nabla_p U \, (F_{\Upsilon}
- {\bar F}_{\Upsilon} ) \right] \:.
\eea
As discussed in subsection 2.1, the scalar fields $Y^I$
($I=0,1,\dots,n$) are determined in terms of an array of $2(n+1)$
harmonic functions $(H^I, H_I)$, given in (\ref{612}), through the
so-called generalised stabilisation equations \cite{9801081,0009234},
\begin{displaymath}
\left(\begin{array}{c}
Y^I - {\bar Y}^I \\
F_I(Y, \Upsilon) - {\bar F}_I ({\bar Y}, {\bar \Upsilon})
\end{array} \right)
=i
\left(
\begin{array}{c}
H^I \\
H_I
\end{array} \right) \;.
\end{displaymath}
For a static solution, $H^I \nabla_p H_I - H_I \nabla_p H^I = 0 $
and $\Upsilon = {\bar \Upsilon} = -64 (\nabla_p U)^2$.

For a holomorphic function of the form (\ref{prepo}) we have
\cite{9906094}
\bea
 i \left[ {\bar Y}^I \, F_I (Y, \Upsilon)
- {\bar F}_I ({\bar Y}, {\bar \Upsilon}) \, Y^I \right]
&=& (S + \bar S) \, \left( \ft12 \, H_m^2
- 128 c_1 \, (U')^2 \right)\;, \\
128 i \, {\rm e}^U \,
\nabla^p \left[ {\rm e}^{-U} \, \nabla_p U \, (F_{\Upsilon}
- {\bar F}_{\Upsilon} ) \right] &=& 128 c_1
\left[ (S + {\bar S})  \left( (U')^2 - U''
- \frac{2}{r} U' \right) - (S + {\bar S})' \,U' \right] \;, \nonumber
\eea
where $U' = dU/dr$ and $(S + {\bar S})' = d (S + {\bar S})/dr$.  By combining
these expressions we obtain
\bea
{\rm e}^{-2U} = \ft12 \, (S + \bar S) \, H_m^2 -
128 c_1 \left[ (S + {\bar S})  \left( U'' +
\frac{2}{r} U' \right) + (S + {\bar S})' \, U' \right] \;.
\label{formu}
\eea
The real part of the dilaton field $S$, on the other hand, is
determined by \cite{9906094}
\bea
S + {\bar S} = 2 \, \sqrt{\frac{H_e^2 \, H_m^2 - (H_e \cdot H_m)^2}{
H_m^2 \left[ H_m^2 - 512 c_1 (U')^2 \right]}} \;,
\label{dila}
\eea
where we have introduced the target-space duality invariant combinations
\bea
H_e^2 &=& 2 \left( - H_0 \, H^1 + \ft14 H_a \eta^{ab} H_b \right) \;,
\nonumber\\
H_m^2 &=& 2 \left( H^0 \, H_1 + H^a \eta_{ab} H^b \right) \;,
\nonumber\\
H_e \cdot H_m &=& H_0 \, H^0 - H_1 \, H^1 + H_2 H^2 + \dots H_n \, H^n
\;.
\eea
Note that in the duality basis of perturbative heterotic string
theory the electric $H$-vector is given by $(H_0, -H^1, H_2,
\dots,H_n)$, whereas the magnetic $H$-vector reads $(H^0, H_1,
H^2, \dots, H^n)$.

By combining (\ref{dila}) and (\ref{formu}) we obtain a non-linear
differential equation for $U$.  Similar non-linear differential equations
have been recently discussed and solved in
\cite{0411255,0411272,0501014}.

Linearising in $c_1$ still yields a complicated differential
equation, namely
\bea
{\rm e}^{-2U} &=& \ft12 \, (S_0 + {\bar S}_0) \, H_m^2 -
128 c_1 \left[ (S_0 + {\bar S}_0)  \left( U'' +
\frac{2}{r} U' - 2 (U')^2 \right) + (S_0 + {\bar S}_0)' \, U' \right]
\nonumber\\
&&+ {\cal O} (c_1^2) \;,
\label{formulin}
\eea
where
\bea
S_0 + {\bar S}_0 = 2 \, \sqrt{\frac{H_e^2 \, H_m^2 - (H_e \cdot H_m)^2}{
H_m^2 \, H_m^2}} \;.
\eea

To utilise the 4D/5D-connection we take the harmonic functions $H^I$
and $H_I$ to be given as in (\ref{129}).
If the charges carried by the solution are generically non-vanishing,
then the solution describes a one-center black hole solution with entropy
given by \cite{9906094}
 \bea {\cal S}_{\rm macro} = - \ft12 \pi (S + \bar S)
\left( p^2 + 512 c_1 \right) \;, \label{entrohet}
 \eea
 where $S$
denotes the value of the dilaton field at the horizon.  This value
is determined by the attractor equations for $S$, which read
\cite{9906094}
 \bea q^2 - |S|^2 p^2 & = & - 2(S+\bar S) (c_1 \, \Upsilon
\, S + {\rm h.c.}) \,,
\nonumber\\
2 i p\cdot q + (S-\bar S) p^2 & = & - 2 (S+\bar S) (c_1 \, \Upsilon
 - {\rm h.c.}) \,, \label{attrs}
  \eea
 where $\Upsilon$ takes the value $-64$ on the horizon. The
combinations $q^2, p^2$ and $q \cdot p$ denote the following
target-space duality invariant combinations of the charges \cite{9906094},
 \bea q^2
&=& 2 q_0 p^1 - \ft12 q_a \eta^{ab} q_b \;,\nonumber\\
p^2 &=& - 2 p^0 q_1 - 2 p^a \eta_{ab} p^b  \;,\nonumber\\
q \cdot p &=& q_0 p^0 - q_1 p^1 + q_2 p^2 + \dots +q_n p^n \;.
 \eea
In the duality basis of perturbative heterotic string
theory the electric charge vector is given by $(q_0, -p^1, q_2,
\dots,q_n)$, whereas the magnetic charge vector reads $(p^0, q_1,
p^2, \dots, p^n)$.

Inserting (\ref{attrs}) into (\ref{entrohet}) yields the entropy
 \bea
  {\cal S}_{\rm macro} &=& \pi \sqrt{q^2 p^2  - (q \cdot p)^2} \,
\sqrt{1 + \frac{512 c_1}{p^2}} \;.
 \label{fivedentrocorr}
 \eea
This describes the $R^2$-corrected entropy of the black hole with
generic charges.  Now consider
restricting the charges to $(q_0 = 2 J, q_A, p^0 = 1, p^A = 0)$
(where $A= 1, \dots,n$).  Note that $p^0 \neq 0$ is a necessary
condition for the 4D/5D-connection \cite{0503217,0504126}.
Then the entropy (\ref{fivedentrocorr})
becomes
 \bea
  {\cal S}_{\rm macro} =2 \pi \sqrt{\ft14 q_1 q_a \eta^{ab} q_b - J^2}
  \, \sqrt{1 - \frac{256
c_1}{q_1}} \;.
  \eea
When $c_1 = 0$, this describes the entropy
of a
charged five-dimensional rotating BPS black hole in a Taub-NUT
geometry.  It is then tempting to conjecture that (\ref{fivedentrocorr})
describes the $R^2$-corrected entropy of the
five-dimensional BPS black hole in a Taub-NUT
geometry.  This is supported by the recent work \cite{0505094}.

In the absence of $R^2$-interactions ($c_1 =0$) the entropy
(\ref{entrohet}) becomes equal to \cite{9507090,9512031}
 \bea
 {\cal S}_{\rm macro} = \sqrt{q^2 \, p^2 - ( q \cdot p)^2 } \;.
 \label{entrohetc1}
 \eea
This follows by inserting (\ref{attrs}) into (\ref{entrohet}).
Inspection of (\ref{entrohetc1}) shows that solutions with charges
satisfying $p^2 = q \cdot p =0$ have zero entropy in the absence of
$R^2$-interactions.  However, in the presence of $R^2$-interactions
the entropy ceases to be vanishing, as can be seen from
(\ref{entrohet}). These solutions therefore provide examples of black
hole solutions which grow a horizon due to $R^2$-interactions, thereby
cloaking the singularity which is present in the absence of higher
curvature interactions.

In the following we will be interested in solutions with $p^0 \neq 0$
so as to be able to utilise the 4D/5D connection. Then, demanding $p^2
= q \cdot p =0, q^2 \neq 0$ results in $q_0 = q_1 = p^a = 0$. The
solutions are therefore allowed to carry non-vanishing electric
charges $(-p^1, q_2 ,\dots, q_n)$.  The interpolating solution
(\ref{formu}) is therefore constructed out of the following
non-trivial harmonic functions
 \bea
 H^0 = n + \frac{p^0 R}{r} \;\;\;,\;\;\; H^1 = h^1 + \frac{p^1 (RG_4)^{1/3}}{r}
\;\;\;,\;\;\; H_a = h_a + \frac{q_a (RG_4)^{2/3}}{R r} \;,
 \eea
whereas the remaining harmonic functions are constant, namely
$H_0 = m, H_1 = h_1, H^a = h^a$.  Note that the constraint
$H^I \nabla_p H_I - H_I \nabla_p H^I = 0 $ results in (here
we set $G_4 = R^2$ )
\bea
h^a \, q_a = m \, p^0 + h_1 \, p^1 \;.
\eea
In the absence of $R^2$-interactions ($c_1 =0)$, the
four-dimensional solution has a naked singularity at $r=0$. This
can be seen from (\ref{formulin}), which then reads
 \bea
 {\rm e}^{- 2 U} = \sqrt{H_e^2 H_m^2 - (H_e \cdot H_m)^2}\;,
 \eea
and which behaves as $r^{-3/2}$ at $r=0$.

In the presence of $R^2$-interactions, however, the solution grows a
horizon. Inspection of (\ref{attrs}) shows that the dilaton then
takes the following value at the horizon,
 \bea
 S + {\bar S} = \sqrt{ \frac{q^2}{- 2 c_1 \Upsilon}} = \sqrt{-
\frac{q_a \eta^{ab} q_b }{256 c_1 }} \;.
 \eea
For this to be a positive quantity, the signs of the charges $q_a$
 have to be chosen in the appropriate way.  The associated
 $R^2$-corrected entropy reads \cite{0412287}
\bea
 {\cal S}_{\rm macro} = - 256 c_1  \pi (S + \bar S)
= \pi \sqrt{ - 256 c_1 \, q_a \eta^{ab} q_b}
 \;. \label{entrohetr2}
 \eea
Thus we see that a three-charge black hole with charges $(p^0, q_2,
q_3)$ in four dimensions (or more generally a black hole with charges
$(p^0, p^1, q_2, \dots, q_n)) $ has a non-vanishing entropy which goes
as $\sqrt{c_1}$, once $R^2$-interactions are taken into account.

Evidence has been presented in \cite{0505094} that the connection
\cite{ 0503217} between five-dimensional BPS solutions in a Taub-NUT
space and four-dimensional BPS solutions continues to hold in the
presence of $R^2$-interactions.  Using this connection, we conclude
that the cloaking of the four-dimensional singularity of the
three-charge solution also takes place in the five-dimensional
solution when taking into account $R^2$-effects.  The cloaking of the
five-dimensional singularity should be such that the entropy of the
resulting five-dimensional two-charge black hole in the Taub-NUT
geometry is given by (\ref{entrohetr2}).

The cloaking of five-dimensional singularities should not only apply
to horizons with $S^3$ topology, but also to horizons with topology
$S^1 \times S^2$, i.e. to black rings.\footnote{This has also been
pointed out and studied in the recent paper \cite{0506215}. Instanton 
corrections may, in principle, also contribute to the cloaking 
\cite{9704095}.}  After
all, using the 4D/5D connection, black ring solutions descend to
multiple center solutions in four dimensions \cite{0504126}.  The
latter may, in the absence of $R^2$-interactions, have multiple naked
singularities which get cloaked by $R^2$-interactions.  This then
implies a cloacking of black ring singularities in five dimensions.
For instance, a two-center solution in four dimensions is connected to
a five-dimensional black ring solution, if one of the centers (say at
$r=0$) carries the entire NUT charge $p^0$, whereas the second center
carries all the other charges \cite{0504126}.  Without
$R^2$-interactions, the second center is a naked singularity if the
charges are restricted to satisfy $p^2 = p \cdot q =0$.  Since $p^0 =
0$ at this center, this implies that $q_1p^1 =0$.  The second center
is therefore allowed to carry electric charges $(q_0, - p^1, q_2,
\dots, q_n)$.  In the presence of $R^2$-interactions the second center
gets cloaked and its entropy is given by \cite{0412287}
\bea
 {\cal S}_{\rm macro} = \pi \sqrt{512 c_1 \, q^2} = 
\pi \sqrt{ 512 c_1 \, (2 q_0 p^1 - \ft12 q_a \eta^{ab} q_b )}
 \;. \label{entrohetbr}
 \eea
 This should describe the entropy of a cloaked black ring in five
 dimensions.  One may also consider other examples, for instance a
 four-dimensional two-center solution where one of the centers carries
 charges $(p^0, q_2, q_3)$, whereas the other center carries charges
 $(q_0, -p^1, q_2, q_3$).  In the absence of $R^2$-interactions these
 two centers describe naked singularities.  Turning on
 $R^2$-interactions should then lead to a cloaking of the two naked
 singularities.  In five dimensions, this would correspond to the
 cloaking of a black hole sitting at the center of a Taub-NUT geometry
 and of a black ring away from the center.

\noindent
{\bf Acknowledgments}

We would like to thank Bernard de Wit, J\"urg K\"appeli, Dieter
L\"ust, Thomas Mohaupt, Kasper Peeters and Tom Taylor for useful
discussions. The work of S.M. is supported by Alexander von Humboldt
Foundation. S.M. would also like to thank Dieter L\"ust and the String
Theory Group at the Arnold Sommerfeld Center LMU for the nice
hospitality during the course of this work.
\noindent


\bigskip



%

\providecommand{\href}[2]{#2}\begingroup\raggedright\endgroup

\end{document}